# High voltage transition metal-free cathode material LiBC$_3$F$_4$ for Li ion batteries

Zhiqiang Wang[1], Lixin Xiong[1], Wenwei Luo[1], Bo Xu[1], Chuying Ouyang[1]*

[1]Department of Physics, Laboratory of Computational Materials Physics, Jiangxi Normal University, Nanchang, 330022, China

*Corresponding author's E-mail: cyouyang@jxnu.edu.cn

**Abstract**

The structural stability and electrochemical performance of boron substituted fluorinated graphite as a Li ion batteries cathode material are studied by first principles calculations. The results show that boron substituted fluorinated graphite BC$_3$F$_4$ possesses excellent structural stability, good electrical and ionic conductivities. Unexpectedly, the average Li intercalation voltage of LiBC$_3$F$_4$ is up to 4.44 V, which is much larger than that of LiBCF$_2$. The average voltage of LiBC$_3$F$_4$ is even larger than that of common commercial transition metal oxides cathodes, indicating that LiBC$_3$F$_4$ is a breakthrough of transition metal-free high voltage cathode materials. By comparing the Fermi level, we found the Fermi level of LiBC$_3$F$_4$ is 1.38 eV lower than that of LiBCF$_2$, leading to the decrease of the electron filling energy for the Li intercalation and forming much higher voltage. Moreover, LiBC$_3$F$_4$ shows small volume expansion and high energy density. LiBC$_3$F$_4$ is a promising high voltage cathode material for Li ion batteries. Finally, by calculating the evolution of Li intercalation voltage and Fermi level during the discharging process, a linear correlation between the Fermi level and Li intercalation voltage has been found.

## 1. Introduction

Li ion batteries (LIBs) have been wildly used in consumer electronics, electric vehicles, large-scale energy storage applications[1-4]. Recently, with the rapid development of electric vehicles and large-scale energy storage, the demands for higher energy density LIBs are more and more urgent. Current commercial cathode materials are transition metals oxides (TMOs), whose capacities are mainly limited by the electrons that TMs can exchange[5]. The capacity of TMOs can be improved via the redox reaction at oxygen sites, leading to the release of oxygen and poor cycle performance[6, 7]. Hence, it is crucial to design new electrode materials with higher capacity and energy density. In our previous, based on fluorinated graphite, new transition metal-free cathode materials B-C-F compounds with ultrahigh capacity and energy density are proposed[8]. Fluorinated graphite CF has been used as a cathode material of Li primary batteries over the past several decades[9]. The large bandgap (5 eV)[8] lead to the occupation of the electrons transferred from Li on the high energy anti-bonding states during the discharging process. Under high Li intercalation concentration, the C-F bonds are broken and to form the LiF compound and graphite. Due to the poor structural stability, CF can not be used as cathode materials for Li ion batteries. By *p*-type doping strategy (substitution carbon by boron atoms), holes are created in the valence-band maximum (VBM), avoiding the occupancy high energy antibonding states of the electrons from Li. Finally, the structural stability and Li intercalation voltage are dramatically improved. The Li intercalation voltage and capacity of new designed cathode material $LiBCF_2$ are 3.49 V and 395.4 mAh/g, respectively, delivering an ultrahigh energy density 1379.9 Wh/Kg[8]. However, compared with carbon, boron is one electron less, leading to form the electron deficiency system. The electron deficient $sp^3$ hybridization mode is not favor of the structural stability and synthesis in experiment. However, the predicted experimental synthesis condition of $LiBCF_2$ is under high temperature and high pressure. In $LiBCF_2$, the ratio of B:C is up to 1:1, can the structural stability be improved by decreasing the B:C ratio? What is the influence of B:C ratio on the electrochemical performance?

In this work, the structural stability and electrochemical performance of boron substituted fluorinated graphite $LiBC_3F_4$ are studied by first principles calculations. By decreasing the boron substitution, the structural stability of $BC_3F_4$ is improved, compared with $BCF_2$ (fully charged state). $LiBC_3F_4$ shows good electrochemical performances as a new cathode material. The average Li intercalation voltage is up to 4.44 V. The capacity and energy density of $LiBC_3F_4$ are 206.75 mAh/g and 917.81 Wh/Kg, respectively. Furthermore, the electrical and ionic conductivities of $LiBC_3F_4$ are studied. By calculated the sequential Li intercalation voltage and Fermi level during the discharging process, a linear relationship between the Fermi level and Li intercalation voltage has been revealed.

## 2. Computational details

All calculations are performed in the Vienna *ab initio* simulation package (VASP) [10]. The projector augmented-wave (PAW) approach[11] has been used to solve Kohn-Sham equations. Based on generalized gradient approximations (GGA), the Perdew–Wang 91 (PW91) functional has been used to describe the exchange-correlation interaction[12]. The kinetic energy cutoff of plane-waves is 500 eV. The Monkhorst-Pack mesh for *k*-point sampling[13] is 3×3×5 for the 2×2×1 supercell. The atomic positions are fully relaxed until the force on every atom is less than 0.02 eV/Å. The effect of van der Waals (vdW) interaction was included by using an empirical correction method proposed by Grimme (DFT-D2)[14]. To analyze the charge transfer between Li and $BC_3F_4$ layer, Bader charge[15] calculations have been performed. The nudged elastic band (NEB) method[16] is used for optimizing the Li ion migration pathways and calculating the energy barriers.

The Li intercalation voltage is defined as

$$V=-[E(Li_mBC_3F_4)-E(Li)-E(Li_nBC_3F_4)]/(m-n)e \quad (0 \leq n < m \leq 1)$$

Where $E(Li_mBC_3F_4)$, $E(Li_nBC_3F_4)$, $E(Li)$, and e are the total energies of the $Li_mBC_3F_4$, $Li_nBC_3F_4$, bulk Li metal, and the charge of one electron, respectively. When *m*=1 and *n*=0, the Li intercalation voltage V is the average voltage of the whole discharging process.

## 3. Results and discussion

### 3.1 Structural evolution of BC$_3$F$_4$ with the Li intercalation

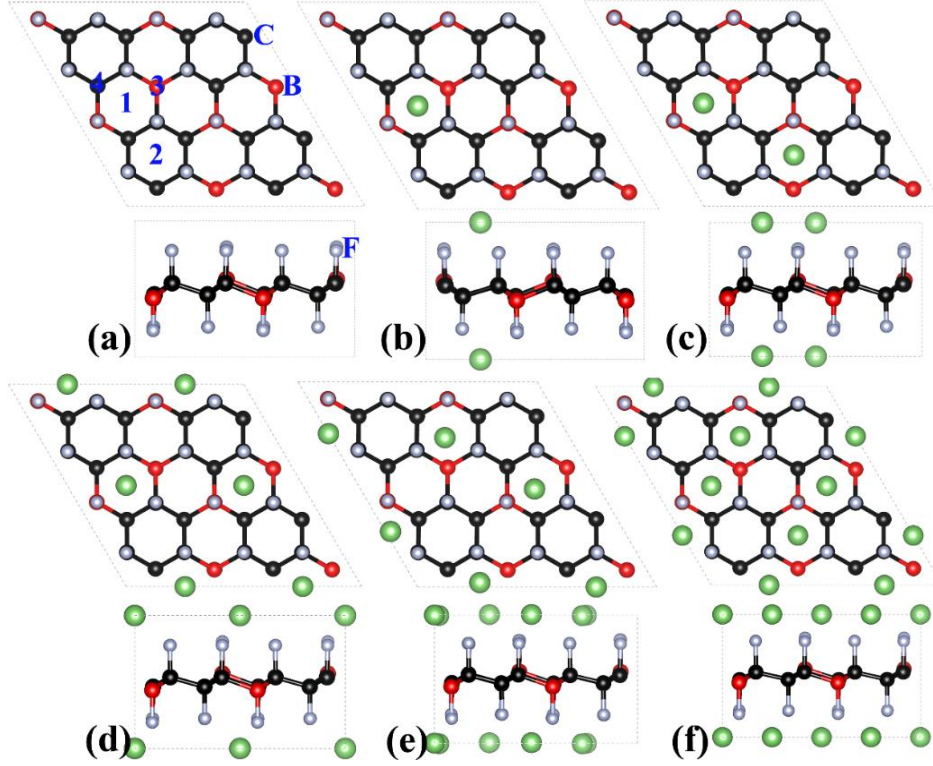

Figure 1. Top and side views of the atomic structures of (a) BC$_3$F$_3$, (b) Li$_{0.125}$BC$_3$F$_4$, (c) Li$_{0.25}$BC$_3$F$_4$, (d) Li$_{0.5}$BC$_3$F$_4$, (e) Li$_{0.75}$BC$_3$F$_4$, (f) LiBC$_3$F$_4$. The big green, small red, black, and light grey ball are Li, B, C, and F atoms, respectively. Four inequitable intercalated sites are marked by the blue numbers.

Table 1. Lattice constants $a$, $c$ (in Å) and C-C, B-C, C-F, B-F bond lengths (in Å) in Li$_x$BC$_3$F$_4$ (0≤$x$≤1).

|  | $a$ (Å) | $c$ (Å) | $b_{C-C}$ (Å) | $b_{B-C}$ (Å) | $b_{C-F}$ (Å) | $b_{B-F}$ (Å) |
| --- | --- | --- | --- | --- | --- | --- |
| BC$_3$F$_4$ | 10.653 | 5.748 | 1.563 | 1.714 | 1.370 | 1.355 |
| Li$_{0.125}$BC$_3$F$_4$ | 10.650 | 5.836 | 1.563 | 1.680 | 1.409 | 1.405 |
| Li$_{0.25}$BC$_3$F$_4$ | 10.660 | 5.848 | 1.564 | 1.668 | 1.418 | 1.411 |
| Li$_{0.50}$BC$_3$F$_4$ | 10.670 | 5.885 | 1.565 | 1.678 | 1.417 | 1.413 |
| Li$_{0.75}$BC$_3$F$_4$ | 10.684 | 5.879 | 1.548 | 1.675 | 1.425 | 1.427 |
| LiBC$_3$F$_4$ | 10.717 | 5.806 | 1.552 | 1.662 | 1.483 | 1.481 |

The top and side views of the atomic structure of BC$_3$F$_4$ are shown in Figure 1(a). BC$_3$F$_4$

is a layered material with the AA stacking. The good dynamic stability has been evaluated by the phonon dispersion as shown in figure S1. No imaginary frequency has been found in the whole Brillouin zone. By comparing the atomic structure of $BCF_2$ and $BC_3F_4$ under 1000 K for the first principles molecular dynamics simulation (figure S2 (a) and S3), it can be found that the structural stability has been improved by decreasing the B/C ratio. The lattice constant and bond lengths are listed in Table 1. The buckling height is 0.8 Å, larger than that of CF and $BCF_2$. The B-F and C-F bond lengths are 1.355 and 1.370 Å, respectively, slightly larger than that in $BCF_2$ (1.343 Å for B-F bond and 1.356 Å for C-F bond). Our results show that the Li intercalation site 1 (the hollow site of the six-membered ring which is consists of two boron and four carbon atoms, marked as the 2B-4C six-membered ring) is the most stable. The Li intercalation voltage is up to 4.87 V, which is an unexpected high voltage for the carbon-based cathode. The mechanism of the high voltage will be discussed later. To study the evolution of the atomic structure and voltage of $BC_3F_4$ with the increasing Li intercalation concentration, the atomic structure and total energy of the $BC_3F_4$ with the different Li intercalation concentrations and Li ion distributions are calculated. The atomic structures of $Li_{0.25}BC_3F_4$, $Li_{0.5}BC_3F_4$, $Li_{0.75}BC_3F_4$, $LiBC_3F_4$ are shown in figure 1(c)-(f). It can be found that the structures are well maintained under high Li intercalation concentration. The lattice constants $a$ and $c$ increase from 10.653 Å and 5.748 Å in $BC_3F_4$ to 10.717 Å and 5.806 Å in $LiBC_3F_4$. Generally speaking, the small expansion of lattice constant facilitates good cycling stability. The lattice constant $a$ increases with the increasing Li intercalation concentration. However, the lattice constant $c$ increases with the increasing Li intercalation concentration when $x \leq 0.5$ in $Li_xBC_3F_4$ and monotonously decrease when $0.5 \leq x \leq 1$. When more and more Li ion are intercalated into the interlayer space, the Coulomb attraction between Li ion and F ion pull the two adjacent B-C-F layers closer and induce the decrease of lattice constant $c$. A similar phenomenon has been reported for the evolution of lattice constant $c$ of $LiCoO_2$ under charging and discharging processes[17]. The B-F and C-F bond lengths increase from 1.355 and 1.370 Å in $BC_3F_4$ to 1.481 and 1.483 Å in $LiBC_3F_4$, respectively. Both the B-F and C-F bonds are well maintained when the Li intercalation concentration is up to 1, which has been confirmed by the first principles molecular dynamics simulation at 1000 K (Figure S2 (b)). During the whole Li intercalation process, the change of lattice constants is very small and the structures are well maintained.

**3.2 Electrochemical performance of $BC_3F_4$ as a cathode material**

Based on the excellent structural stability of $BC_3F_4$ under the Li intercalation, we shift our attention to its electrochemical performance. Li intercalation voltage is a key parameter for cathode materials. As we mentioned above, when the Li concentration is 0.125 (figure 1(b)), the Li intercalation voltage is 4.87 V. For the whole Li intercalation process, the average Li intercalation voltage is 4.44 V, which is much larger than that of $LiBCF_2$ (3.49 V), $LiB_2C_2F_2$ (3.63 V) and common commercial cathode materials. The capacity and energy density of $LiBC_3F_4$ are 206.75 mAh/g and 917.81 W·h/Kg, respectively, which are smaller than that of $LiCoO_2$, but much larger than that of $LiFePO_4$ and $LiMn_2O_4$.

Furthermore, both electrical and ionic conductivities are important for the rate performance of cathode materials. Boron has one less valence electron, compared with carbon, hence, boron substitution at the carbon site is typical *p*-type doping. Holes are generated at the valence band maximum as shown in Figure 3. The electronic structure change from semi-conductive for fluorinated graphite (CF) to metallic for $BC_3F_4$, indicating excellent electrical conductivity at the beginning of the discharging process. With the increasing Li intercalation concentration, more and more holes are filled by the electron transferred from Li to $BC_3F_4$ layers. When the Li intercalation concentration reaches to 1, the electronic structure change to semiconductor with a 2 eV bandgap. The metallic electronic structure for $Li_xBC_3F_4$ (0≤x<1) and a small bandgap for $LiBC_3F_4$ are helpful to guarantee the excellent electrical conductivity of the $Li_xBC_3F_4$ cathode material.

Then, the ionic conductivities under the fully charged and discharged states are studied.

As shown in figure 4 (a) and (b), under the fully charged state, two inequitable Li ion migration pathways have been taken into consideration. The first migration pathway is from the hollow site of the 2B-4C six-membered ring to the nearest neighbor hollow site of the 2B-4C six-membered ring. The second migration pathway is from the hollow site of the 2B-4C six-membered ring to the nearest neighbor hollow site of the six-membered ring which is consists of six carbon atoms. The energy barriers are 0.68 and 0.70 eV, respectively. Under the fully discharged states, the Li ion migration energy barrier is 0.44 eV. The above energy barriers of Li ion migration are slightly lower or higher than that of common cathode materials. Hence, the ionic conductivity of $Li_xBC_3F_4$ cathode material is acceptable.

In general, $LiBC_3F_4$ shows good electrochemical performances as a new transition metal-free cathode material for Li ion batteries. The average Li intercalation voltage is up to 4.44 V, which is larger than that of common commercial cathode materials. The capacity and energy density of $LiBC_3F_4$ are 206.75 mAh/g and 917.81 W·h/Kg, respectively. Excellent electrical conductivity has been confirmed by the calculations of the electronic structure. The volume expansion during the Li deintercalation process is only 1.9%, which is much smaller than that of $LiCoO_2$, facilitating good cycling stability. Importantly, $LiBC_3F_4$ is a new transition metal-free high voltage cathode material, avoiding the consumption of high-cost Co and Ni elements.

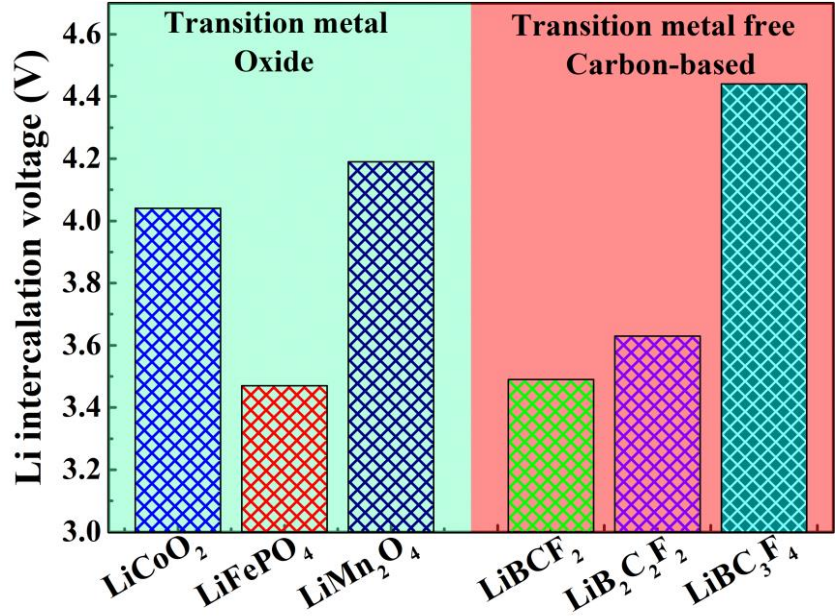

Figure 2. The average Li intercalation voltage of the B-C-F system, compared with common commercial cathode materials of Li ion batteries.

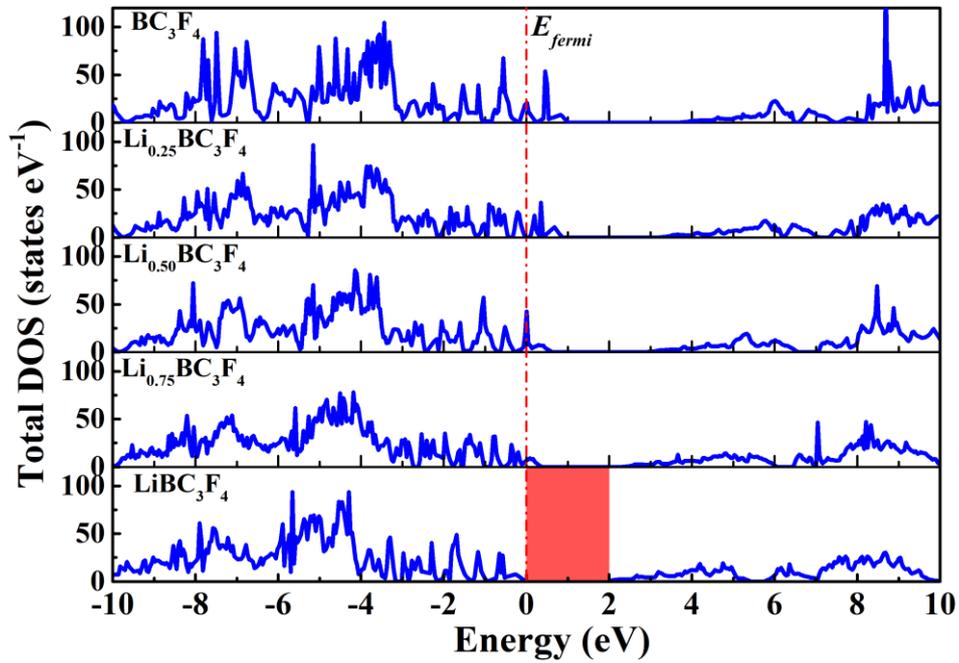

Figure 3. Total density of states of Li$_x$BC$_3$F$_4$ (0≤$x$≤1).

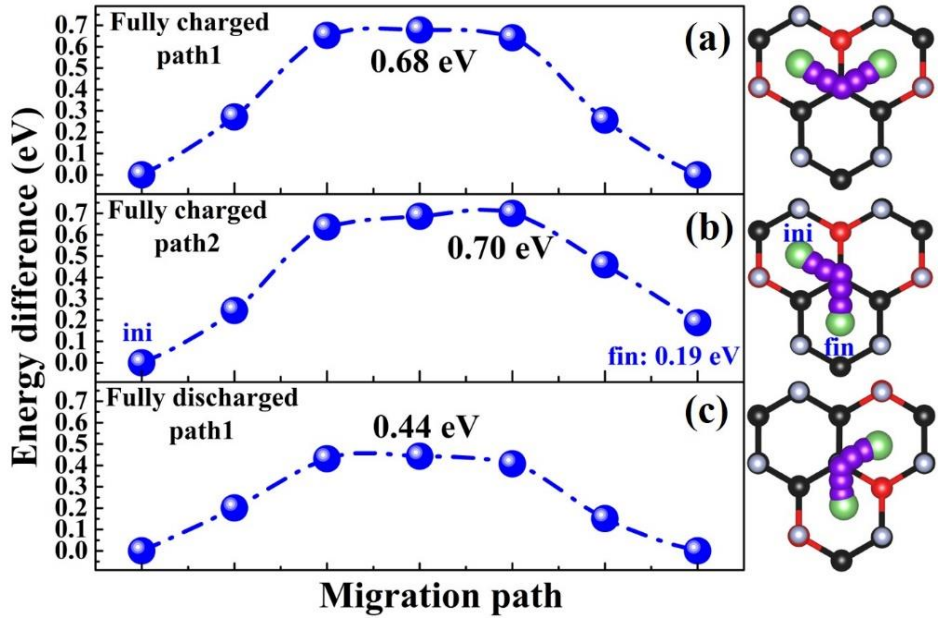

Figure 4. The energy barrier of Li ion migration in $BC_3F_4$ (fully charged state) and $LiBC_3F_4$ (fully discharged state). The atomic structures of the Li ion migration paths are showed.

### 3.3 Correlation between Fermi level and Li intercalation voltage

Finally, we want to give a short discussion about the correlation between the Fermi level and Li intercalation voltage. As mentioned above, the average Li intercalation voltage of $LiBCF_2$ and $LiBC_3F_4$ are 3.49 and 4.44 V, respectively. To figure out the mechanism, we calculated the Fermi level of $LiBCF_2$ and $LiBC_3F_4$, the results show that the Fermi level of $LiBC_3F_4$ is 1.38 eV lower than that of $LiBCF_2$. The remarkable decrease of the Fermi level leads to the much smaller electron filling energy for the electron from Li, leading to achieving much higher Li intercalation voltage. To exhibit the correlation between the Fermi level and Li intercalation voltage more clearly. We calculated the sequential Li intercalation voltage during the whole Li intercalation process. As shown in figure 5(b), the sequential voltage slightly decreased when $x<0.5$, and the Fermi level slightly increased. However, when $x=0.5$, there is a sudden decrease (0.66 V) of the sequential voltage. Correspondingly, there is a sudden increase (0.63 eV) of the Fermi level. Then, the sequential voltage slightly decreased when $0.5<x<1$, and the Fermi level slightly increased. However, when $x=1$, there is a sudden decrease (2.12 V) of the sequential voltage. Correspondingly, there is a sudden increase (2.37 eV) of the Fermi level. When $x=1$, the electronic structure of $LiBC_3F_4$ is a semiconductor with a 2 eV bandgap. When more lithium are intercalated into $LiBC_3F_4$, No holes can accommodate the electron transferred from lithium, the corresponding electron has to occupy the high energy orbit (conduction band minimum) (as shown in figure 5(a)). Hence, the electron filling energy is dramatically increased, leading to the decreased of the sequential Li intercalation voltage. Generally speaking, the lower the Fermi level, the higher the Li intercalation voltage. Furthermore, for $BC_3F_4$, the Li intercalation concentration can not larger than the boron substitution concentration, due to the semi-conductive electronic structure of $LiBC_3F_4$. Our results show a clear linear relationship between Fermi level and Li intercalation voltage (as shown in Figure 5(c)) which is helping to design new anode and cathode materials for metal ion secondary batteries.

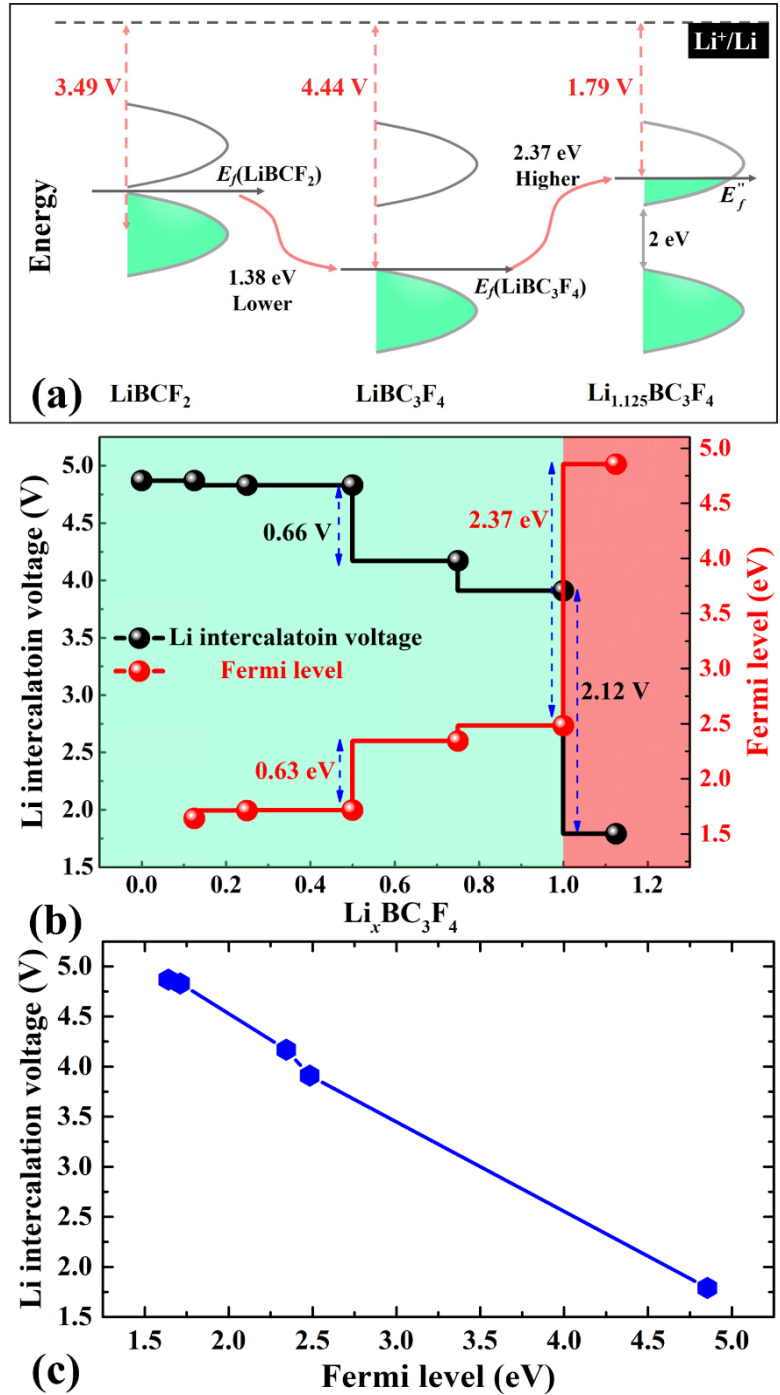

Figure 5. (a) Illustration of the relationship between Fermi level Li intercalation voltage for B-C-F system. (b) The sequential Li intercalation voltage and Fermi level of $Li_xBC_3F_4$ (0≤x≤1.125). (c) The sequential Li intercalation voltage, a function of Fermi level of $Li_xBC_3F_4$ system.

## 4. Conclusions

In summary, the structural stability and electrochemical performance of boron substituted fluorinated graphite as a cathode material of Li ion batteries are studied by first principles calculations. By decreasing the boron concentration, the structural stability of $BC_3F_4$ is improved. $LiBC_3F_4$ shows excellent electrochemical performances as a new cathode material. The average Li intercalation voltage is up to 4.44 V. The capacity and energy density of $LiBC_3F_4$ are 206.75 mAh/g and 917.81 W·h/Kg, respectively. both the good electrical and ionic conductivities are. The volume expansion during the Li deintercalation process is very small (1.9%). $LiBC_3F_4$ is a promising high voltage cathode material for Li ion batteries. What's more, $LiBC_3F_4$ is a transition metal-free high voltage cathode material.

Furthermore, we discussed the correlation between the Fermi level and Li intercalation voltage. $LiBC_3F_4$ shows a much higher Li intercalation voltage (4.44 V) than $LiBCF_2$ (3.49 V). The average voltage of $LiBC_3F_4$ is even larger than that of common commercial transition metal oxides cathodes, indicating that $LiBC_3F_4$ is a breakthrough of transition metal-free high voltage cathode materials. By comparing the Fermi level, we found the Fermi level of $LiBC_3F_4$ is 1.38 eV lower than that of $LiBCF_2$, leading to the decrease of the electron filling energy for the Li intercalation and forming much higher voltage. Finally, by calculating the evolution of Li intercalation voltage and Fermi level during the discharging process, a linear correlation between the Fermi level and Li intercalation voltage has been found. With the increasing Li intercalation concentration, the Fermi level gradually increases and leads to the decrease of Li intercalation voltage. An additional conclusion that the Li intercalation concentration can not larger than the boron substitution concentration, due to the semi-conductive electronic structure of $LiBC_3F_4$ can be drawn.

## Acknowledgments

This work is supported by the Natural Science Foundation of China (NSFC) under Grant No. 51962010. The computations were partly performed on TianHe-2(A) at the National Supercomputer Center in Tianjin.

# Supporting information

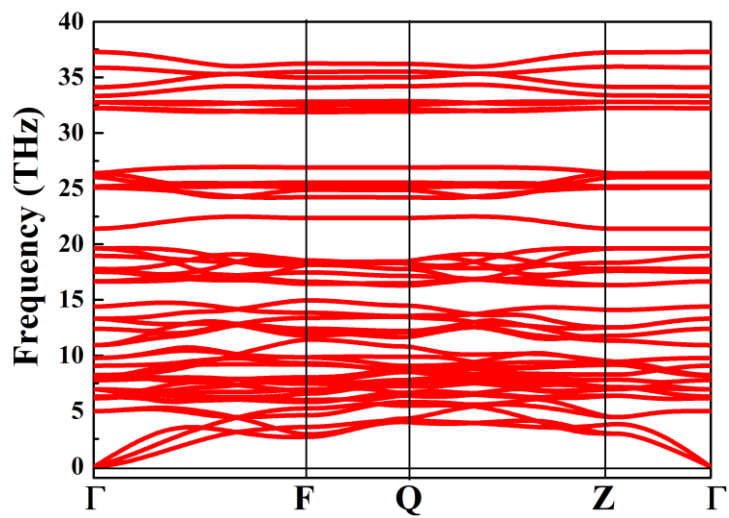

Figure S1. The phonon dispersion of BC$_3$F$_4$.

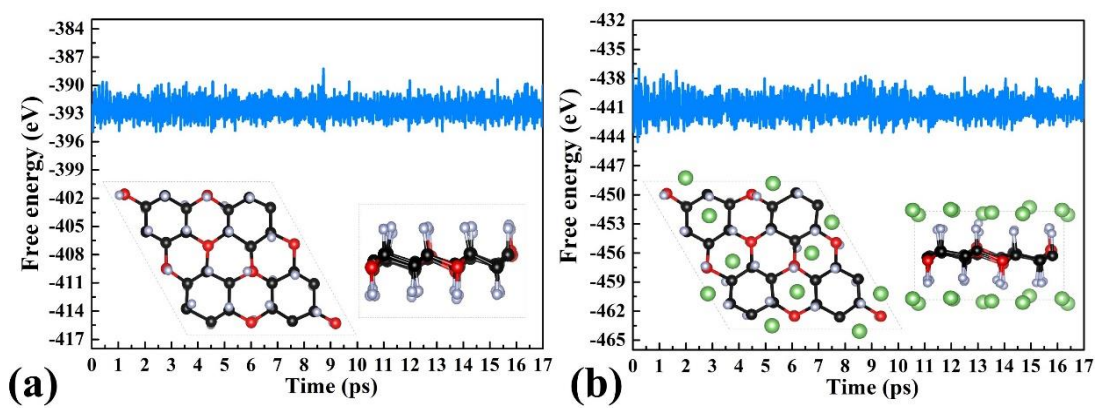

Figure S2. Variation of free energy of (a) BC$_3$F$_4$ and (b) LiBC$_3$F$_4$ during ab initio molecular dynamic simulation at a temperature of 1000 K. the illustration are the atomic structures of BC$_3$F$_4$ and LiBC$_3$F$_4$ at t=17 ps.

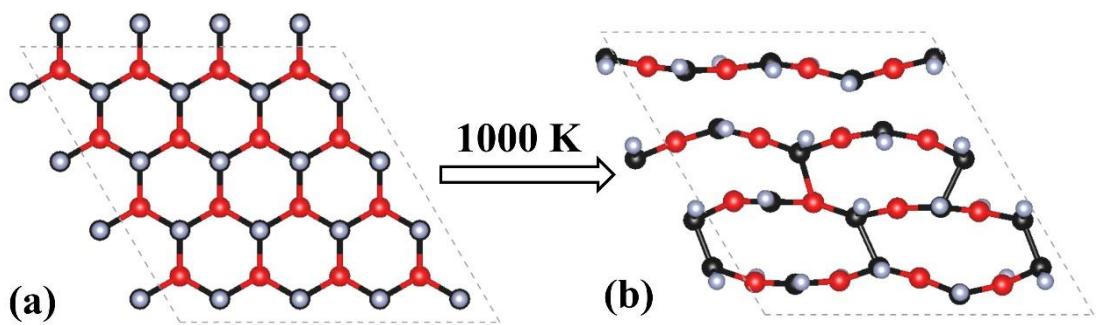

Figure S3. The top view of the atomic structures of BCF$_2$ (a) 0 K and (b) at 1000 K for the *ab* initio molecular dynamics simulation.